# Unique Physically Anchored Cryptographic Theoretical Calculation of the Fine-Structure Constant α Matching both the g/2 and Interferometric High-Precision Measurements

Charles Kirkham Rhodes

Department of Physics, University of Illinois at Chicago,
Chicago, IL 60607-7059, *USA*

Ultrabeam Technologies, LLC
Chicago, IL 60611

## ABSTRACT

The fine-structure constant α, the dimensionless number that represents the strength of electromagnetic coupling in the limit of sufficiently low energy interactions, is the crucial fundamental physical parameter that governs a nearly limitless range of phenomena involving the interaction of radiation with materials. Ideally, the apparatus of physical theory should be competent to provide a calculational procedure that yields a quantitatively correct value for α and the physical basis for its computation. This study presents the first demonstration of an observationally anchored theoretical procedure that predicts a unique value for α that stands in full agreement with the best (~370 ppt) high-precision experimental determinations. In a directly connected cryptographic computation, the method that gives these results also yields the magnitude of the cosmological constant $\Omega_\Lambda$ in conformance with the observational data and the condition of perfect flatness ($\Omega_\Lambda + \Omega_m = 1.0$). Connecting quantitatively the colossal with the tiny by exact statements, these findings testify that the universe is a system of such astonishing perfection that an epistemological limit is unavoidably encountered.

## BACKGROUND

In his book [1], *QED: The Strange Theory of Light and Matter*, Richard P. Feynman described the fine-structure constant *α*, introduced initially by Arnold Sommerfeld into the realm of physics in 1916 through the analysis of relativistic corrections to the Bohr atom, as "one of the greatest damn mysteries of physics: a magic number that comes to us with no understanding by man." Physical theory should be competent to provide a specific procedure that yields both (1) a quantitatively correct value for α and (2) the physical basis for its computation. In order to be fully capable of describing complex phenomena quantitatively, this theoretical system should possess certain characteristics. They are (i) the capacity for high complexity, (ii) sharp precision (scale-free), (iii) a clear organizing principle, (iv) a physically defined unique basis, and (v) the ability to formulate an optimization. It has been shown [2] that a properly defined Galois field readily satisfies these five requirements. A cryptographic system based on a finite field [3,4] $\mathbb{F}_P$ with a prime modulus $P$ fundamentally represents a precisely organized procedure of counting that possesses an associated cyclic group of units $\mathbb{F}_P^*$ containing $P-1$ elements. Therefore, under the condition that the prime $P$ is physically anchored, the cryptosystem is uniquely defined, utterly devoid of any free parameters, and possesses the limiting exactitude of unity regardless of the magnitude of the prime $P$ designating the field order or the complexity of the group structure legislated by the divisors of the integer ($P-1$).

In principle, physical theory can be represented with a procedure of counting without compromise or limitation, as first declared in 1896 by Ernst Mach [5], "Jede mathematische Aufgabe könnte durch direktes Zählen gelöst werden," and echoed a year later by David Hilbert in his famous *Zahlbericht* [6]. This study presents the first demonstration of a theoretical procedure, grounded physically by precise quantitative agreement with observational data [7] encompassing the six intrinsic universal parameters α, G, h, c, $\Omega_\Lambda$, $\Omega_m$, and perfect flatness $\Omega_\Lambda$ +$\Omega_m$=1.0, and mathematically defined by a prime $P_\alpha \equiv 1 \pmod 4$ that is very sharply constrained by an independent array of physically motivated arithmetic stipulations and whose size (~$6.759 \times 10^{60}$) corresponds exactly to the Planck mass $(\hbar c/G)^{1/2}$, that (a) uniquely yields the magnitude of $\alpha^{-1}$ = 137.0359991047437444154 in agreement with both of the recently measured values [8,9] ($\alpha^{-1}$ = 137.035999084(51) and $\alpha^{-1}$ =137.03599945(62)), (b) predicts a convergence value of $\alpha$ for future experimental determinations to a level better than 1 part in $10^{20}$, and (c) gives a coherent ensemble of predictions that can be tested in future experiments. They are the electron neutrino mass $m_{\nu_e}$ = 0.8019 meV, the muon neutrino mass $m_{\nu_\mu}$ = 27.45 meV, the masses of a superheavy (>$10^{18}$ GeV) supersymmetric Higgs pair, and the value of the unified strong-electroweak coupling constant $\alpha^* = (34.26)^{-1}$.

An essential element of this analysis is the independent demonstration [7] that the solution of the Higgs Congruence $B_{Higgs}^2 \equiv -1 \pmod{P_\alpha^2}$ in the extension field $\mathbb{F}_{P_\alpha^2}$ yields the cosmic parameters $\Omega_\Lambda$ and $\Omega_m$ in full accord with their experimental determinations [7] and the relation $\Omega_\Lambda + \Omega_m = 1.0$, the condition of perfect flatness. These findings provide sharp additional constraints on the cryptographic analysis and consequently provide strong confirming evidence for the correctness of the selected magnitude [7] for the prime modulus $P_\alpha$ discussed below.

Finally, since the summation of the results obtained by these methods [2,7,9] demonstrates the existence of precise quantitative relationships that connect governing cosmic-scale entities ($G$, $\Omega_\Lambda$, and $\Omega_m$) to fundamental micro-scale quantities ($\alpha^{-1}, m_{\nu_e},$ and $m_{\nu_\mu}$), a profound conclusion inescapably follows, specifically, that the universe represents a system of astonishing cosmic perfection whose level of regulation is one part in ~ $P_\alpha^2 \approx 10^{121}$. An impenetrable epistemological limit is thereby unavoidably encountered, since no measurement whatsoever can ever approach this stupendous precision.

## METHODS

The chief plan of this work is the development of the maximal set of physically meaningful and quantitatively precise statements connecting particle states and interactions that can be formulated with this cryptographic picture. Four parallel primary consequences follow; (α) the theoretical analysis receives the broadest possible foundation, (β) the volume of associated constraints, both interlocking the physical entities with the mathematical structure of the cryptographic apparatus and testing its internal consistency, is maximized, (γ) a unique value for the fine-structure constant α is derived, and (δ) a platform is erected from which predictions of several key physical parameters are launched. Indeed, the subsequent computation of $\Omega_\Lambda$ and $\Omega_m$ cited above [7] has already demonstrated the capacity to extend directly the theoretical approach yielding $\alpha$ to the quantitative description of other fundamental physical quantities and we foresee the continuation of this extension.

The ability to compute a quantitatively correct value for the fine-structure constant $\alpha$ that has independent physical and mathematical bases requires the introduction of several fresh concepts. In conjunction, these ideas define a new organizing principle for the description of physical properties and interactions that is founded on <u>precise physically anchored modular counting</u>. In essence, it is found that the intrinsic structure of a properly constructed and equipped cryptographic system utilizing a finite field naturally presents a mathematical template so fitting that it can be considered as the ideal correspondence for the classification of physical particle states and their interactions. The key act enabling the realization of this concept is the identification of the <u>unique</u> prime modulus $P_\alpha$, established in detail below, that defines the finite field $\mathbb{F}_{P_\alpha}$ upon which all computations rest. Since the extension field $\mathbb{F}_{P_\alpha^2}$ is also fully defined by $P_\alpha$, the successful computation of $\Omega_\Lambda$ and $\Omega_m$ as a manifestation of the Higgs state [7], <u>without the incorporation of any additional parameters</u>, further corroborates the selected magnitude of $P_\alpha$ by providing a coherent theoretical synthesis quantitatively relating the six intrinsic universal parameters $\alpha$, G, h, c, $\Omega_\Lambda$, and $\Omega_m$ that stands in full conformance with the corresponding extant observational data, including the constraint $\Omega_\Lambda + \Omega_m = 1.0$ for perfect flatness.

## RESULTS

### Précis of Established Correspondences

As a fundamental element, the theoretical description involves the association of the masses of particles with properly normalized integer magnitudes $B$ (representatives of residue classes $[B]_{P_\alpha}$) in the field $\mathbb{F}_{P_\alpha}$ in which the prime modulus $P_\alpha$ is uniquely defined by both observational physical data and a set of independent mathematical requirements. These mass numbers, as shown in Table I, aside from their customarily considered magnitudes, possess several additional mathematical properties that sharply enhance and extend their theoretical descriptive power. This analysis attributes physical significance to all aspects of their mathematical endowment. As represented by the correspondences shown in Table I, the précis of prior studies illustrated in Table II, and the key particle identifications and mathematical relations presented in Table III, previous work has produced a substantial body of findings that is subsumed in the following analysis.

Specifically, these earlier studies have demonstrated that (a) the parity [2,11] of $B$ distinguishes fermi and bose particle species, (b) the inverse $[B]_{P_\alpha}^{-1}$ of the mass number $[B]_{P_\alpha}$ defines a new physical state [2] that perforce satisfies $[B]_{P_\alpha}[B]_{P_\alpha}^{-1} \equiv 1 \pmod{P_\alpha}$, (c) the set of divisors $\{d_B\}$ of the mass number $B$ specifies informational content designating particle attributes through the introduction of the concept of "genetic divisors" [12], (d) these genetic divisors have a quantitative physical measure of genetic comparison though the construction of a physically motivated metric [13] expressed with *p*-adic numbers [14,15], (e) the subgroup orders $\{\delta_{P_\alpha-1}\}$ of $\mathbb{F}^*_{P_\alpha}$ given by the set $\{d_{P_\alpha-1}\}$ of divisors [3] of the integer $P_\alpha-1$, are the basic classifiers of particle states and interactions [2,16], (f) the concept of supersymmetric fermi/bose particle pairing $[B/B_{ss}]$, together with the specific quantitative mass relationship governing the pair given by

$$[B]_{P_\alpha} + [B_{ss}]_{P_\alpha} \equiv 0 (\mod P_\alpha), \tag{1}$$

finds natural incorporation in the definition of particle states and their associated group properties [2,13,16], (g) the first supplementary law of Quadratic Reciprocity [17], the "aureum theorema" of Gauss that served as the foundation of modern number theory, is equivalent to the statement of the symmetry $[B^2_{\text{Higgs}} \equiv -1 (\mod P_\alpha)]$ defining the Higgs system [18], (h) the prospective $\nu_e$ and $\nu_\mu$ mass numbers [2], respectively designated by two even primitive roots $g_\alpha$ and $g_\beta^{-1}$ of $P_\alpha$, are divisors of the integer $P_\alpha-1$ and obey the seesaw congruence

$$g_\alpha g_\beta^{-1} \equiv B^2_{\text{Higgs}} (\mod P_\alpha) \equiv -1 (\mod P_\alpha), \tag{2}$$

(i) the Bézout identity [2], under the physically based condition $\gcd(g_\alpha, g_\beta^{-1}) = 2$, both relates the $\nu_e$ and $\nu_\mu$ systems to their corresponding supersymmetric and inverse state counterparts and partitions $\mathbb{F}^*_{P_\alpha}$ by optimally establishing two maximally disjoint sets of subgroups with respective orders $g_\alpha$ and $g_\beta^{-1}$, and (j) the utterly crucial result that the computation of the Higgs state [7] in the extension field $\mathbb{F}_{P_\alpha^2}$ yields the cosmic parameters $\Omega_\Lambda$ and $\Omega_m$ in agreement with all present observational data and the condition of exact flatness $\Omega_\Lambda + \Omega_m = 1.0$.

We note additionally, that the appropriate equivalence relation used for modular computations with residue classes in a finite field, as illustrated with several entries given in Tables I, II and III, is the congruence, a generalized mathematical statement of equality originated by Carl F. Gauss [19]. Furthermore, since the cryptographic procedures developed below reduce simply to (a) counting precisely with (b) large numbers, we observe that these developments simultaneously answer the query of Wigner concerning the basis of the efficacy of mathematics for the description of physical phenomena [20] with the first characteristic and resolve the mystery of the significance of large magnitudes posed by Dirac [21,23], Eddington [24], and others [25] with the second; organized counting with large integers connects the large and the small with exact statements.

**Table I: Mathematical Properties and Physical Correspondences of Mass Number Integers (B)**

| Mathematical Property | Physical Correspondence | Corresponding Mathematical Statement | Remarks | References |
|---|---|---|---|---|
| **Parity** | Fermion / $B$ even<br>Boson / $B$ odd | $B \equiv 0$ (mod 2) Fermi<br>$B \equiv 1$ (mod 2) Bose | Fermion / Boson parity choice legislated by supersymmetry and the conservation of angular momentum in two-body decay processes with the identification of the state of the electron neutrino $\nu_e$ with the even primitive root mass number $g_\alpha$. | 2,11 |
| **Magnitude** | $B$ integer, particle mass number proportional to physical mass $m_B$. | $B$ is a representative of the residue class $[B]_{P_\alpha}$ in the field of $\mathbb{F}_{P_\alpha}$. | Physical mass is given by $m_B = BE_o$ with energy unit $E_o = (\hbar c^5 / G)^{1/2} / P_\alpha = 1.8062 \times 10^{-33}\, eV$. See Table II. | 2,3,7,26-28 |
| **Factor Stucture/ Divisor Set** | $B = \prod_{i=1}^{n} P_i^{\alpha_i}$, $\{d_B\}$ | Number of divisors of mass number $B$ given by $d(B) = \prod_{i=1}^{n}(\alpha_i + 1).$ | For smooth propagation, generally require $d(B^2) = \prod_{i=1}^{n}(2\alpha_i + 1)$ large to enable a high multiplicity of solutions corresponding to kinetic motion for a particle of mass m through the normalized relativistic energy ($E$)/momentum ($p$) equation $E^2 - p^2 = m^2$. Genetic divisor concept applies to divisor set $\{d_B\}$. p-adic metric endows the genetic divisors with a quantitative comparison of different particle states that distinguishes their properties. See Table II. | 2,11,13,29,30 |

**Table I: Mathematical Properties and Physical Correspondences of Mass Number Integers (B) (con't)**

| | | | | |
|---|---|---|---|---|
| **Order $\delta_B$** | Subgroup order establishes families of similar particles and designates interactions. For the $\nu_e$ and $\nu_\mu$ neutrinos, they have the maximal order $\delta = P_\alpha - 1$. This indicates status as a primitive root that acts mathematically as a generator of the field $\mathbb{F}_{P_\alpha}$ and physically labels flavor transforming propagation. For the subset of primitive roots of $P_\alpha$ that are divisors of $P_\alpha - 1$, the integers must be even, designating Fermi charater. For the Higgs state, $\delta_{Higgs} = 4$. Generally, for $\delta_B$ even, the supersymmetric particle mass number $[B_{ss}]_{P_\alpha}$ has the same order as $[B]_{P_\alpha}$; conversely, for $\delta_B$ odd, the subgroup containing $[B]_{P_\alpha}$ does not contain $[B_{ss}]_{P_\alpha}$. | $B^{\delta_B} \equiv 1 \pmod{P_\alpha}$. The set of subgroup orders $\{\delta_{P_\alpha}\}$ of the cyclic group $\mathbb{F}^*_{P_\alpha}$, such that $\delta_B \in \{\delta_{P_\alpha}\}$, is given by the set of divisors $\{d_{P_\alpha - 1}\}$ of $P_\alpha - 1$, hence, $\delta_B \in \{d_{P_\alpha - 1}\}$. | Genetic divisor concept and p-adic metric also apply to subgroup orders. Supersymmetry is a combined particle and group property, since particle $P$ and its supersymmetric partner $P_{ss}$ are members of the same subgroup only for even orders. See Table II. | 2,3,11-18,32 |
| **Inverse** | $[B]^{-1}_{P_\alpha}$ corresponds to a physical particle state $P_{in}$ that is a member of the canonical octet illustrated in Fig. (1). | $[B]^{-1}_{P_\alpha} \equiv [B]^{P_\alpha - 2}_{P_\alpha} \pmod{P_\alpha}$ | Inverse relation follows from Fermat's Little Theorem. If $B$ is a divisor of $P_\alpha - 1$, $[B_{ss}]^{-1}_{P_\alpha}$ is given directly by Lemma 1 as $(P_\alpha - 1)/B$. See entry for the Bézout Identity in Table II. | 2, 26-28,31 |

**Table II: Summary of Prior Results**

| FINDING | MATHEMATICAL STATEMENT | PHYSICAL INTERPRETATION | REMARKS | REFERENCES |
|---|---|---|---|---|
| **Supersymmetric Mass Relationship** | $[B]_{P_\alpha} + [B_{ss}]_{P_\alpha} \equiv 0 \,(\text{mod}\, P_a)$ | $m + m_{ss} = k m_p, \; k = 1,2,3,...$ Sum of particle $P$ mass ($m$) and supersymmetric particle $P_{ss}$ mass ($m_{ss}$) equals a multiple of the modular mass given by the Planck mass $m_p = (\hbar c / G)^{\frac{1}{2}}$. See Table III. | With $P_\alpha \equiv 1 \,(\text{mod}\, 4)$, since the Euler totient function $\varphi(P_\alpha - 1)$ is always even for any prime $P > 3$, and a result of Gauss certifies that all primitive roots can be arranged in supersymmetric pairs as $g_\alpha + g_\beta = P_\alpha$, we have the supersymmetric mass relationship shown in Fig.(1) and stated by Eq.(2). This condition is adopted for all supersymmetric mass pairs, an outcome confirmed by the subgroup structure of $\mathbb{F}_{P_\alpha}^*$; with $P_\alpha \equiv 1 \,(\text{mod}\, 4)$, all even subgroups contain perforce both $[B]_{P_\alpha}$ and $[B_{ss}]_{P_\alpha}$. | 2,7,11,16,29,30 |
| **Supersymmeteric Higgs Pair** | $B_{Higgs}^2 \equiv -1 \,(\text{mod}\, P_a)$ | Since the symmetry condition defining the Higgs state is $B_{ss} = B_{in}$, the quartet pattern shown in Fig.(1) collapses to a doublet of states such that $B_{Higgs}^{boson} + B_{Higgs}^{fermion} \equiv 0 \,(\text{mod}\, P_\alpha)$. See Table III. | The Higgs Congruence has a solution if and only if $P_\alpha \equiv 1(\text{mod}\, 4)$ and is equivalent to the first supplementary law of Quadratic Reciprocity. The cryptographic solution for Higgs mass solves exactly, with no approximation, a traditionally many-body problem in quantum field theory in a single step. | 16,18, 27-29 |
| **Higgs Seesaw Congruence** | $[g_\alpha]_{P_\alpha} [g_\beta]_{P_\alpha}^{-1} \equiv$ $[B_{Higgs}]_{P_\alpha}^2 \,(\text{mod}\, P_\alpha) \equiv$ $-1(\text{mod}\, P_\alpha)$ | The integer $g_\alpha$ represents the electron neutrino mass with $m_{\nu_e} \cong 0.8019 \; meV$; $g_\alpha$ is a primitive root of $P_\alpha$, status that designates flavor changing propagation. The integer $g_\beta^{-1}$, mutatis mutandis, gives the muon neutrino mass with $m_{\nu_\mu} \cong 27.45 \; meV$. See Fig. (1). | Since $g_\alpha$ and $g_\beta^{-1}$ are primitive roots of $P_\alpha$, they are generators of the field $\mathbb{F}_{P_\alpha}$. From the Higgs Congruence, we also have $[g_\alpha]_{P_\alpha} [g_\beta]_{P_\alpha}^{-1} \equiv -1(\text{mod}\, P_a)$. This congruence is an equivalence relation linking the concepts of mass (left) and space (right). See Table III. | 2,18 |

**Table II: Summary of Prior Results (con't.)**

| | | | | |
|---|---|---|---|---|
| **Universe Mass Number** | $E_u = P_\alpha^2 + 1$<br><br>$P_\alpha$ also gives $\Omega_\Lambda$ and $\Omega_m$ from the Super Higgs Congruence $B_{Higgs}^2 \equiv -1 \pmod{P_\alpha^2}$. | $G = (P_\alpha^2 + 1)\hbar c / M_u^2 = 6.67 \times 10^{-8} \, cm^3/gs^2$ in conformance with the observed value. $E_u$ is even and represents a fermion. Ultimately, $P_\alpha$ is anchored to the intrinsic universal constants $\Omega_\Lambda$ and $\Omega_m$. | Explicit p-adic form of $E_u$ represents residue class $[1]_{P_\alpha}$. Value of $P_\alpha$ determined by Higgs symmetry in the extension field $\mathbb{F}_{P_\alpha^2}$. | 2,7,13-15 |
| **Supersymmetric Particle of the Universe Mass Number** | $(E_u)_{SS} = 2P_\alpha - 1$, boson, $2P_\alpha - 1$, prime. | The minimum value of $(E_u)_{SS}$ corresponds to $2P_\alpha - 1$. Specific numerical test discloses that the integer $2P_\alpha - 1$ is a prime. | Explicit p-adic form of $(E_u)_{SS}$ represents residue class $[-1]_{P_\alpha}$. A boson with mass $2P_\alpha - 1$ is stable against all decay modes, from gamma burst analysis. With $2P_\alpha - 1$ prime, a field $\mathbb{F}_{2P_\alpha - 1}$ is defined such that the set of subgroup orders $\{\delta_{2P_\alpha - 1}\}$ of the cyclic group $\mathbb{F}_{2P_\alpha - 1}^*$ contains the set of subgroup orders of $\mathbb{F}_{P_\alpha}^*$, since $(P_\alpha - 1)$ divides $(2P_\alpha - 2) = 2(P_\alpha - 1)$. | 2,7,11 |
| **Energy Unit $E_o$** | $E_o = [1]_{P_\alpha}$ | $E_o = (\hbar c^5 / G)^{1/2} / P_\alpha = 1.8062 \times 10^{-33} \, eV$. | Derived on the basis of supersymmetry and the process of gravitational renormalization in the computation of $\Omega_\Lambda$ and $\Omega_m$ in the extension field $\mathbb{F}_{P_\alpha^2}$. Gives mass of $\nu_1$ particle presented in Table III. | 7 |
| **Fine-Structure Constant $\alpha$** | $\alpha^{-1} = \dfrac{4g_\beta^{-1}}{g_\alpha} =$<br>137.0359991047437444154.<br>Integers $g_\alpha$ and $g_\beta^{-1}$ are even primitive roots of $P_\alpha$. | $\alpha^{-1} = \dfrac{4m_{\nu_\mu}}{m_{\nu_e}}$ | Relates $\alpha$ directly to the ratio of the masses of two key fermions, the electron $\nu_e$ and muon $\nu_\mu$ neutrinos, whose mass numbers $g_\alpha$ and $g_\beta^{-1}$ are even primitive roots of $P_\alpha$ and divisors of $P_\alpha - 1$. See Eq. (2), Eq. (4), and Fig. (2). | 2,18 |

**Table II: Summary of Prior Results (con't.)**

| | | | | |
|---|---|---|---|---|
| **Unified Strong-Electroweak Coupling Constant** $\alpha^*$ | $(\alpha^*)^{-1} = \dfrac{g_\beta^{-1}}{g_\alpha}$ | $(\alpha^*)^{-1} = \dfrac{m_{\nu_\mu}}{m_{\nu_e}} = \dfrac{\alpha^{-1}}{4} \cong 34.26$ | Predicted value falls within expected range for this parameter. Determined by $\nu_e / \nu_\mu$ neutrino (fermion) mass ratio. | 2,18,31 |
| **Genetic Divisors** | $B = \prod_{i=1}^{n} P_i^{\alpha_i}$ gives set of divisors $\{d_B\}$ whose number is $d(B) = \prod_{i=1}^{n}(\alpha_i + 1)$. | Individual divisors in the set of divisors $\{d_B\}$ interpreted as informational elements describing particle attributes. An example is the divisor 2 that identifies all fermions (e.g. even values for $g_\alpha$ and $g_\beta^{-1}$ corresponding to the $\nu_e$ and $\nu_\mu$ neutrinos, respectively). See Fig. (1). | Concept applicable for both mass numbers $B$ and subgroup orders $\delta_B \in \{d_{P_\alpha-1}\}$ and is rendered quantitative with the p-adic metric. See Table I. | 2,12,26,28 |
| **p-adic Metric** | For two particles with mass numbers x and y, the genetic distance is $D(x,y) = \dfrac{1}{\gcd(x,y)}$. | On the rational number field $\mathbb{Q}$, the metric distance $D(x,y)$ between a pair of unequal mass particles x and y is given by the inverse of the greatest common divisor (gcd) of their respective mass numbers. Similar particles (x,y), systems that possess a large $\gcd(x,y)$, perforce have a small value for $D(x,y)$. This outcome has an obvious physical motivation; namely, particles with closely related physical properties and a large corresponding genetic similarity enjoy a small metric separation. For two fermi species (x,y), since they are associated with even mass numbers, the maximum value is $D(x,y) = \frac{1}{2}$. Since bose particles (x,y) have mass numbers that are odd, the corresponding maximal value for either a bose/bose or a fermi/bose pair is $D(x,y) = 1$. | Metric distance $D(x,y)$ applicable to both particle mass numbers $B$ and subgroup orders $\delta_B \in \{d_{P_\alpha-1}\}$. See Table I. | 2,12,13 |

**Table II: Summary of Prior Results (con't.)**

| | | | | |
|---|---|---|---|---|
| **Bézout Identity** | Bézout Identity states that for any two integers $x, y$, $rx + sy = \gcd(x, y)$ for suitable integer values r and s. Overall, $\gcd(g_\alpha, g_\beta^{-1}) = 2$ and $g_\alpha$ and $g_\beta^{-1}$ both are divisors of $P_\alpha - 1$ such that $g_\alpha g_\beta^{-1} = P_\alpha - 1$. | *Lemma (1):* Let $P_\alpha$ be prime, $B$ be a divisor of $P_\alpha$-1, and $(B)_{ss} = P_\alpha - B.$ Then, $(B)_{ss}^{-1} = (P_\alpha - 1)/B$. Hence, $g_\alpha g_\beta^{-1} = P_\alpha - 1$, with $\gcd(g_\alpha, g_\beta^{-1}) = 2$. From the properties of inverse states and the Bézout identity, we have $g_\alpha g_\alpha^{-1} + g_\beta g_\beta^{-1} \equiv \gcd(g_\alpha, g_\beta^{-1}) \equiv$ $[D(g_\alpha, g_\beta^{-1})]^{-1} \pmod{P_\alpha}$, a statement relating all four states illustrated in Fig.(1). | With $\gcd(g_\alpha, g_\beta^{-1}) = 2$, $\nu_e$ and $\nu_\mu$ are maximally different particles with $D(g_\alpha, g_\beta^{-1}) = \frac{1}{2}$. Bézout identity supplements multiplicative Seesaw Congruence $g_\alpha g_\beta^{-1} \equiv B_{Higgs}^2 \pmod{P_\alpha} \equiv$ -1 $\pmod{P_\alpha}$ and profoundly influences the subgroup structure of $\mathbb{F}_{P_\alpha}^*$. | 2,18, 27,32 |

**Particle State Definition**

The cardinal issue for the organization of the mass scale is the definition of particle states $P$ in a form that naturally matches the modular language and group structure of a finite field. Since the definition of a group demands that each element $\{x\}$ of the group has an inverse element $\{x^{-1}\}$, as shown in Table I, a new physical entity, the corresponding inverse state $P_{in}$ exists. Two vital consequences follow [2]. They are, as presented in lines 2 and 3 of Table II, (α) the ability to define the Higgs system [33,34] as the fundamental symmetry point in the group structure of the field [2,18,28] in a manner that quantitatively expresses supersymmetric fermion/boson pairing [2], as stated by Eq.(1), and (β) the concepts of mass [35] and space [36] become conjoined by an equivalence relationship [2], the basic seesaw congruence given by Eq.(2).

For the specific particle state definition, we recognize four prominent desired characteristics; they are (A) the equivalence of particle $P$ and its corresponding antiparticle $\bar{P}$ masses ($m_P = m_{\bar{P}}$), a demand of CPT invariance[37,38], (B) the general expression of supersymmetric [39] fermion (f) /boson (b) pairing, (C) the existence of a supersymmetric Higgs state, the entity that introduces mass into the Standard Model [34,38] and (D) a basis for the values of both the fine-structure constant α and the unified strong-electroweak coupling constant α*, the fundamental physical parameters regulating non-gravitational interactions [40].

Taken together, the two features (A) and (B) of the mass spectrum incorporate three forms of two-particle associations. CPT separately pairs both fermi (e.g. $e^+/e^-$) and bose (e.g. $\pi^+/\pi^-$ and $\pi^0/\pi^0$) species. Supersymmetry connects the fermi and bose particle genres through supersymmetric pairs ($P/P_{ss}$). These two classes of relationships produce three pairings (f/f, b/b, and b/f) yielding a system in which each particle $P$ has both an antiparticle $\bar{P}$ and a supersymmetric partner $P_{ss}$. The introduction of the inverse state required by the group properties additionally associates each particle state $P$ with a corresponding inverse state $P_{in}$, again through a precise and specific relationship for the particle masses. The overall result can be represented by an affiliation of four particles $(P, P_{ss}, P_{in}, (P_{in})_{ss})$, normally, but not necessarily, with nondegenerate mass values, that are further related to a corresponding quartet of antiparticles. The final outcome is the formation of an octet of quantitatively related states that generally represents four distinct mass values.

The results of earlier studies [2,28] have demonstrated how this general pattern can represent the electron neutrino $\nu_e$, the muon neutrino $\nu_\mu$, and the Higgs supersymmetric pair, together with specific mass relationships among the particles $P$, their supersymmetric counterparts $P_{ss}$, and the corresponding inverse states $P_{in}$. As stated in Table I, with the consideration of two-body decay amplitudes (e.g. $\pi \to \mu^+ + \nu_\mu$) it was also determined [2,11,28], on the combined basis of supersymmetric classification and the conservation of angular momentum, that the mass numbers of fermions and bosons are naturally partitioned through the parity of the mass integer $B$ by respectively corresponding to even and odd numbers in the field. As a specific example, Figure (1) illustrates this set of relationships for the states that correspond to the electron neutrino $\nu_e$, the muon neutrino $\nu_\mu$, and their associated supersymmetric and inverse state partners [2,28]. Table III presents additional key particle identifications and their corresponding mathematical relationships.

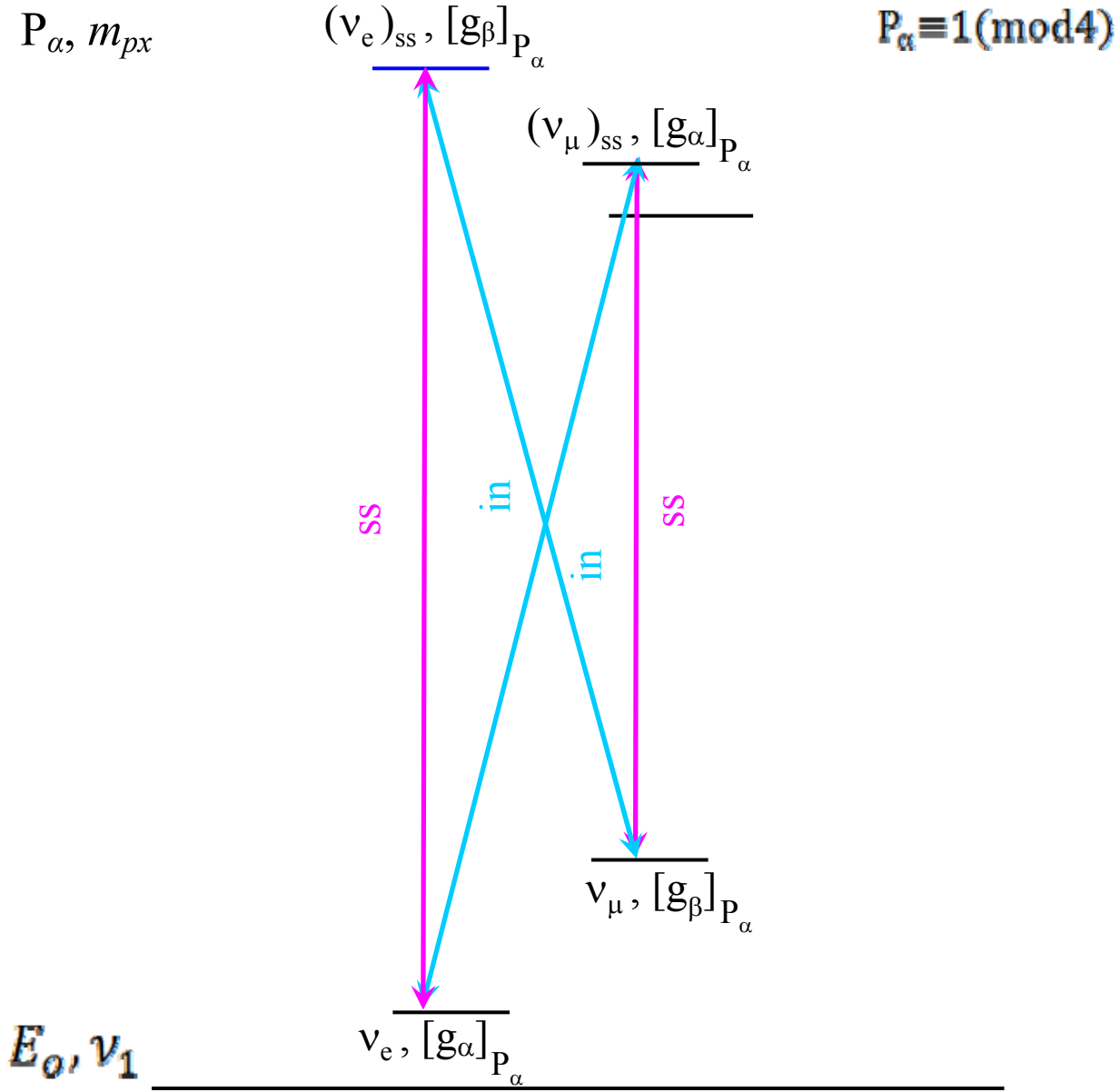

Fig. (1): Illustration of the general state pattern for the mass scale octet with the example of the electron $v_e$ and muon $v_\mu$ neutrinos [2,28]. The supersymmetric (ss) and inverse (in) relationships are shown. The general supersymmetric mass condition given by Eq.(1) is written in terms of residue classes is $[B]_{P_\alpha} + [B_{ss}]_{P_\alpha} \equiv 0 \pmod{P_\alpha}$. With the $v_e$ and $v_\mu$ mass numbers designated respectively by the integers $g_\alpha$ and $g_\beta^{-1}$, representatives of the corresponding residue classes $[g_\alpha]_{P_\alpha}$ and $[g_\beta]_{P_\alpha}^{-1}$, we have $g_\alpha + g_\beta = P_\alpha$ and $g_\alpha^{-1} + g_\beta^{-1} = P_\alpha$. From Lemma 1 in Table II, we obtain $g_\alpha g_\beta^{-1} = P_\alpha - 1$, as stated by Eq.(2), or equivalently, $[g_\alpha]_{P_\alpha} [g_\beta]_{P_\alpha}^{-1} \equiv -1 \pmod{P_\alpha}$. The mass numbers $g_\alpha$, $g_\beta$, $g_\alpha^{-1}$ and $g_\beta^{-1}$ are all primitive roots [27,29] of $P_\alpha$. Mathematical status as a primitive root is associated with the physical manifestation of the flavor changing propagation that is observed for the $v_e$ and $v_\mu$ neutrinos [2,51-53]. On the basis of the prior work [2] represented in Table III, the magnitude of $P_\alpha$ represents the monopole with the Planck mass $m_{px}$, $E_o$ denotes the energy unit, and the $v_1$ is a boson neutrino [11] with a physical mass equal to $E_o$ and mass number unity. Further details are contained in Table II and Table III.

**Table III:** Key Particle Identifications and Mathematical Relationships

| Particle(s) | Mass | Mass Number | Corresponding Mathematical Statement | Remarks | References |
|---|---|---|---|---|---|
| **Magnetic Monopole** | Planck Mass $\sqrt{\dfrac{\hbar c}{G}} \cong 1.22 \times 10^{19}\, GeV$ | $P_\alpha$, modulus of $\mathbb{F}_{P_\alpha} \cdot P_\alpha^2$, modulus of extension field $\mathbb{F}_{P_\alpha^2}$. | $P_\alpha$ determined by $\Omega_\Lambda$ and $\Omega_m$ through Super Higgs Congruence $B_{Higgs}^2 \equiv -1 (\mod P_\alpha^2)$. $G = \dfrac{(P_\alpha^2 + 1)\hbar c}{M_u^2}$, $\left(\dfrac{M_u^2}{\hbar}\right)\left(\dfrac{G}{c}\right) \equiv 1 (\mod P_\alpha)$. | Magnetic Monopole corresponds to Planck mass. Magnitude of $P_\alpha$ specified by Eq. (6) and ultimately anchored by conformance of the Higgs solutions for $\Omega_\Lambda$ and $\Omega_m$ with observational data. $M_u$, $G$, $\hbar$, and $c$ are related to a precision of one part in $\sim 10^{121}$. Value of $G$ is in accord with the experimental value. Product of $(M_u^2/\hbar)$ and $(G/c)$ represents the residue class $[1]_{P_\alpha}$. | 2,7 |
| **Electron Neutrino $\nu_e$** | $m_{\nu_e} = 0.8019\, meV$ | Mass number $g_\alpha$ is an even primitive root of $P_\alpha$. | $g_\alpha g_\beta^{-1} = (P_\alpha - 1) \equiv -1 (\mod P_\alpha)$, $\dfrac{g_\alpha}{g_\beta^{-1}} = \alpha^* = 4\alpha = \dfrac{m_{\nu_e}}{m_{\nu_\mu}} \cong \dfrac{1}{34.26}$. | Along with $g_\beta^{-1}$, $g_\alpha$ is strongly constrained by measured value of $\alpha$ in connection with the divisor structure of $P_\alpha$-1. See Fig. (2), Table II, and Conclusions. | 2,28 |
| **Muon Neutrino $\nu_\mu$** | $m_{\nu_\mu} = 27.45\, meV$ | Mass number $g_\beta^{-1}$ is an even primitive root of $P_\alpha$. | $g_\alpha g_\beta^{-1} \equiv B_{Higgs}^2 (\mod P_\alpha)$, $\gcd(g_\alpha, g_\beta^{-1}) = 2$, $\alpha = \dfrac{g_\alpha}{4 g_\beta^{-1}} = \dfrac{\alpha^*}{4}$. | Mass relationships governing $\nu_e$ and $\nu_\mu$ neutrinos are constrained by the multiplicative seesaw congruence given by Eq. (2) and the additive Bézout identity stated in Eq. (10). The latter influences the group structure of $\mathbb{F}_{P_\alpha}^*$. | 2,11,16,28, 31 |
| **Super-symmetric Higgs Pair** | $B_{Higgs}^f = 2.64 \times 10^{18}\, GeV$ $B_{Higgs}^b = 9.56 \times 10^{18}\, GeV$ | $B_f \cong 1.46 \times 10^{60}$ $B_b \cong 5.29 \times 10^{60}$ | $B_{Higgs}^2 \equiv -1 (\mod P_\alpha)$. Order $\delta_{Higgs} = 4$ in group structure. Computed directly by $g^{\frac{P_\alpha - 1}{4}} \equiv B_{Higgs} (\mod P_\alpha)$ for any primitive root $g$ of $P_\alpha$. | Higgs Congruence is identical to the first supplementary law of Quadratic Reciprocity and defines a point of symmetry in $\mathbb{F}_{P_\alpha}^*$. See Table II. The Higgs concept can be transferred to the extension field $\mathbb{F}_{P_\alpha^2}$. | 2,7,16-18 |
| **Boson Neutrino $\nu_1$** | $m_{\nu_1} \cong 1.8062 \times 10^{-33}\, eV$ | $B_{\nu_1} = 1$, boson particle state that corresponds to energy unit $E_o$. | $E_o = (\hbar c^5 / G)^{1/2} / P_\alpha$, energy unit. Derivable from supersymmetry and the process of gravitational renormalization. See Table II. | The boson neutrino $\nu_1$ state defines the energy unit $E_o$, carries momentum zero, and possesses a maximum energy of unity, its rest mass. See Table II. | 2,7,11 |

The cosmic/micro-scale connection noted at the outset is an intrinsic property of the cryptographic system that is directly embedded in the definition of the particle state $P$. This coupling occurs, since a state $P$ represented by an integer $B$ in a residue class $[B]_{P_\alpha}$, by the definition of a group, necessarily demands the existence of a corresponding inverse state $P_{in}$ denoted by $[B]_{P_\alpha}^{-1}$. The cosmic/micro-scale coupling is made explicit by the general form of the inverse $[B]_{P_\alpha}^{-1}$ given by

$$[B]_{P_\alpha}^{-1} \equiv [B]_{P_\alpha}^{P_\alpha - 2} (\mathrm{mod}\, P_\alpha), \tag{3}$$

a relation that follows directly from Fermat's Little Theorem [27,32]. The operation represented by Eq. (3) manifestly shows that an inverse $[B]_{P_\alpha}^{-1}$ generally depends on <u>both</u> the value of $[B]_{P_\alpha}$ and the cosmically based modulus $P_\alpha$. Hence, the dependence of the inverse on the modulus $P_\alpha$ ensures that <u>all</u> particle states that are defined by the general octet pattern, illustrated by the example of the neutrinos in Fig.(1), have an explicit cosmic connection. Indeed, the mass number $[B]_{P_\alpha}$ of any particle can be properly considered as its cosmic address.

### Uniqueness of the Cryptographic Prime Modulus $P_\alpha$

The validity of the choice of the prime $P_\alpha$ is the central issue in this analysis, as this integer governs the quantitative integrity of all findings. The value of the integer $P_\alpha$ defining the field $\mathbb{F}_{P_\alpha}$ and its uniqueness are established by a septet of constraints founded on both a set of observational data and an independent array of mathematical stipulations. They are (1) the magnitude [2], estimated originally [2] on the measured values of $H, M_\mu$, and $G$, but is <u>ultimately</u> [7] <u>anchored quantitatively</u> to the measured values of $\Omega_\Lambda$, $\Omega_m$, and $G$ under the constraint $\Omega_\Lambda + \Omega_m = 1.0$ by a relationship that gives a normalized value corresponding exactly to the Planck mass $(\hbar c/G)^{1/2}$, (2) the multiply required demand [2,16,28] to be a prime $P_\alpha \equiv 1 (\mathrm{mod}\, 4)$, (3) the optimization [2] of the group structure of $\mathbb{F}_{P_\alpha}^*$ such that maximal complexity is achieved under the double constraint of (i) a bounded magnitude for $P_\alpha$ and (ii) the satisfaction of a physically based condition on the divisor structure of $P_\alpha - 1$ involving the Bézout identity [2,9], conditions enabling the number of divisors of $P_\alpha - 1$ specified by the arithmetic function [26] d($P_\alpha - 1$) given in Table I to be close to the maximum possible for highly composite numbers (HCN) [41-43], (4) the requirement stated in Table II that $2P_\alpha - 1$ be a prime, an outcome legislated by supersymmetry and the definition of the energy unit $E_0$ that enables the condition $\{\delta_{2P_\alpha - 1}\} \supset \{\delta_{P_\alpha}\}$, involving the sets of subgroup orders $\{\delta_{2P_\alpha - 1}\}$ and $\{\delta_{P_\alpha}\}$ of the cyclic groups $\mathbb{F}_{2P_\alpha - 1}^*$ and $\mathbb{F}_{P_\alpha}^*$ given respectively by the divisors of $2P_\alpha - 2$ and $P_\alpha - 1$, to hold, (5) the exceptionally sharp test on the divisors of $P_\alpha - 1$ demanding the existence of a unique ratio $\alpha_2^{-1}$ of the two even primitive roots $g_\alpha$ and $g_\beta^{-1}$ of $P_\alpha$ representing the $\nu_e$ and $\nu_\mu$ states depicted in Fig. (1) that (a) agrees with the measured value of the fine-structure constant $\alpha^{-1}$ in accordance with the relation [2]

$$\alpha^{-1} = \frac{4g_\beta^{-1}}{g_\alpha} \tag{4}$$

to within 370 ppt, and (b) whose product simultaneously satisfies the seesaw congruence $g_\alpha g_\beta^{-1} \equiv P_\alpha - 1 \equiv -1 \pmod{P_\alpha}$ stated by Eq. (2), (6) a second test on the divisors of $P_\alpha - 1$ that requires the two primitive roots $g_\alpha$ and $g_\beta^{-1}$ yielding $\alpha^{-1}$ from Eq. (4) to correspond prospectively to physical mass values for the electron $\nu_e$ and muon $\nu_\mu$ neutrinos that are consistent with current experimentally determined limits, and (7) a third test on the two primitive roots $g_\alpha$ and $g_\beta^{-1}$ that requires $\gcd(g_\alpha, g_\beta^{-1}) = 2$, a condition that (a) constructs a physically significant additive relationship through the Bézout identity that optimally connects the election and muon neutrino masses with their supersymmetric and inverse state counterparts and (b) profoundly influences the subgroup structure [2] of $\mathbb{F}_{P_\alpha}^*$ by establishing two maximally disjoint sets of subgroups comprised respectively of the subgroups with orders $g_\alpha$ and $g_\beta^{-1}$. A summary of these seven constraints is presented in Table IV.

Overall, this concordance of constraints <u>quantitatively</u> tests the magnitudes, prime status, parities, orders, and divisor structures of the integers representing the physical quantities against corresponding experimental data. Consequently, as detailed below and elsewhere [7], within the experimentally determined range established by α, G, h, c, $\Omega_\Lambda$, and $\Omega_m$, the choice of the prime $P_\alpha$ is powerfully limited by a substantial set of physical measurements and an independent host of strict mathematical requirements.

The key step [2] in the determination of the magnitude of $P_\alpha$ is the formation of the divisor structure of $P_\alpha - 1$ in a manner that optimizes the complexity of the group $\mathbb{F}_{P_\alpha}^*$ under the condition that the modulus $P_\alpha$ (a) has an observationally bounded magnitude, (b) respects the conditions on the primitive root divisors $g_\alpha$ and $g_\beta^{-1}$ of $P_\alpha - 1$ stated in Table II involving $\alpha$ and the Bézout identity, and (c) unavoidably gives $P_\alpha \equiv 1 \pmod{4}$. This objective requires the number of divisors of $P_\alpha - 1$ be maximized and that the divisor 4 be among them. Therefore, in the prime factorization of $P_\alpha - 1$, the optimum value of the power $\alpha_2$ of the prime 2 ideally corresponds to the minimal choice

$$\alpha_2 = 2, \tag{5}$$

since this selection simultaneously provides for both the presence of the divisor 4 and satisfaction of the constraint $\gcd(g_\alpha, g_\beta^{-1}) = 2$ presented by the Bézout identity in Table II. This requirement is based on physical grounds by the parity (even) necessary for the two primitive root divisors $g_\alpha$ and $g_\beta^{-1}$ of $P_\alpha - 1$ to represent the fermion states $v_e$ and $v_\mu$. Mathematically, the even parity of both $g_\alpha$ and $g_\beta^{-1}$ is likewise imperative [2] for them simultaneously to be primitive roots of $P_\alpha$ and divisors of $P_\alpha - 1$.

**Table IV: Mathematical and Physical Constraints Determining the Prime Modulus $P_\alpha$**

| INTEGER/ PROPERTY | MATHEMATICAL CONSTRAINTS | PHYSICAL CONSTRAINTS | REMARKS |
|---|---|---|---|
| $P_\alpha$ Magnitude | $B_{Higgs}^2 \equiv -1 (\mod P_\alpha^2)$ in the extension field $\mathbb{F}_{P_\alpha^2}$. | Magnitude bounded by the mass of the universe and quantitatively locked to the measured values of $\Omega_\Lambda$, $\Omega_m$ and G with the constraint $\Omega_\Lambda + \Omega_m = 1.0$. | $B_{Higgs}^2$ solution represents the subgroup of order 4 in $\mathbb{F}_{P_\alpha^2}$ that expresses the Higgs symmetry. The solution pair matches the observational data on $\Omega_\Lambda$ and $\Omega_m$. Supersymetry gives $\Omega_\Lambda + \Omega_m = 1.0$. |
| $P_\alpha$ Magnitude | $P_\alpha \equiv 1 (\mod 4)$ prime. Serves as the modulus defining $\mathbb{F}_{P_\alpha}$. | $P_\alpha$ magnitude corresponds to Planck mass $(\hbar c/G)^{1/2}$. | $P_\alpha \equiv 1 (\mod 4)$ allows solution of Higgs Congruence $B_{Higgs}^2 \equiv -1 (\mod P_\alpha)$ in $\mathbb{F}_{P_\alpha}$. |
| $P_\alpha - 1$ Divisor Structure/ Divisor Order | $g_\alpha g_\beta^{-1} = P_\alpha - 1$; $g_\alpha, g_\beta^{-1}$ even primitive roots of $P_\alpha$. Multiplicative statement involving the respective mass numbers of the $\nu_e$ and $\nu_\mu$ neutrinos. | $\alpha^{-1} = \dfrac{4 g_\beta^{-1}}{g_\alpha}$. $g_\alpha$ is the $\nu_e$ mass number and $g_\beta^{-1}$ is the $\nu_\mu$ mass number. Primitive root status of $g_\alpha$ and $g_\beta^{-1}$ is the mathematical signature associated with the observed flavor changing propagation of neutrinos. | Unique pair of primitive root divisors of $P_\alpha - 1$, $g_\alpha$ and $g_\beta^{-1}$ give $\alpha^{-1}$ matching high precision measurement (~370 ppt). |
| $P_\alpha - 1$ Divisor Structure | <u>Bézout Identity</u> $\gcd(g_\alpha, g_\beta^{-1}) = 2$, minimal optimized value. $[g_\alpha]_{P_\alpha} [(g_\beta)_{ss}]_{P_\alpha} + [g_\beta]_{P_\alpha}^{-1} [(g_\alpha)_{ss}]_{P_\alpha} \equiv 2 (\mod P_\alpha)$ | Relates $\nu_e$ and $\nu_\mu$ particles together with their supersymmetric partners $(\nu_e)_{ss}$ and $(\nu_\mu)_{ss}$ with a single additive statement involving their corresponding mass numbers $(g_\alpha)_{ss}$ and $(g_\beta^{-1})_{ss}$. | The Bézout Identity condition, with the Seesaw Relation $g_\alpha g_\beta^{-1} \equiv B_{Higgs}^2 (\mod P_\alpha) \equiv -1 (\mod P_\alpha)$, influences the group structure of $\mathbb{F}_{P_\alpha}^*$ and connects the $\nu_e$ and $\nu_\mu$ neutrino states to both the Higgs state and Quadratic Reciprocity. |
| $P_\alpha - 1$ Divisor Structure | Arithmetic function $d(P_\alpha - 1)$ maximized within constraints of bounded magnitude of $P_\alpha$ and the condition imposed on the divisor structure of $P_\alpha - 1$ by the Bézout Identity. | Through maximization of the number of subgroups of $\mathbb{F}_{P_\alpha}^*$, the complexity of the classification of physical particle states is likewise maximized. | The function $d(P_\alpha - 1) = \sim 2.3 \times 10^{11}$ has a magnitude close to the maximum possible for highly composite numbers (HCN). |
| $2P_\alpha - 1$ Magnitude | $2P_\alpha - 1$ prime. Enables definition of a field $\mathbb{F}_{2P_\alpha - 1}$. | State with mass number $2P_\alpha - 1$ represents supersymmetric state of the universe. Motivates gravitational renormalization. | Gamma burst analysis shows that all states with mass numbers $\leq 2P_\alpha - 1$ are stable. |
| $2P_\alpha - 1$ Divisor Structure | $2P_\alpha - 1$ prime enables the condition $\{\delta_{2P_\alpha - 1}\} \supset \{\delta_{P_\alpha}\}$, involving the sets of subgroup orders $\{\delta_{2P_\alpha - 1}\}$ and $\{\delta_{P_\alpha}\}$ of the cyclic groups $\mathbb{F}_{2P_\alpha - 1}^*$ and $\mathbb{F}_{P_\alpha}^*$ given respectively by the divisors of $2P_\alpha - 2$ and $P_\alpha - 1$, to hold. | Legislated by supersymmetry and the definition of the energy unit $E_0$. Leads to the concept of gravitational renormalization and the result $E_0 = m_{px} c^2 / P_\alpha = (\hbar c^5/G)^{1/2} / P_\alpha$ $= 1.8062 \times 10^{-33}$ ev.. | Connects the energy unit $E_0$ to the Planck mass through the concepts of supersymmetry and gravitational renormalization. |

The specific candidate for $P_\alpha - 1$ that emerged from the application of this logic [2] is the exceptionally regular 151-smooth high k-factorable integer

$$P_\alpha - 1 = 2^2 \cdot 3^2 \cdot 5^2 \cdot 7 \cdot 11 \cdot 13 \cdot 17 \cdot 19 \cdot 23 \cdot 29 \cdot 31 \cdot 37 \cdot 41 \cdot 43 \cdot 47 \cdot 53 \cdot 59 \cdot 61 \cdot 67 \cdot 71 \cdot 73 \cdot$$
$$79 \cdot 83 \cdot 89 \cdot 97 \cdot 101 \cdot 103 \cdot 107 \cdot 109 \cdot 113 \cdot 127 \cdot 131 \cdot 137 \cdot 139 \cdot 149 \cdot 151 =$$
$$675958604975493531986670708686976004052055490087069632275730 0, \qquad (6)$$

a number that contains 39 factors and 36 distinct primes. All primes from 2 to 151 are represented and the factor structure of $P_\alpha - 1$ is *optimized for information capacity (complexity) under the constraint of an experimentally bounded magnitude for $P_\alpha$, as noted above.* This feature of the design of $P_\alpha - 1$ is revealed in the extra (quadratic) powers of the prime factors 2, 3 and 5, a property that enables an increase in the function $d(P_\alpha - 1)$ and adds minimally to the size of $P_\alpha$ until the product of the additional factors surpasses the next prime (157). Note that $2 \times 3 \times 5 = 30 < 157$, but $2 \times 3 \times 5 \times 7 = 210 > 157$, so this criterion would lead to termination of the quadratic powers with the prime 5, in agreement with the form given in Eq.(6). On the basis of the theory of highly composite numbers [41-43], if $\alpha_2 = 2$, it follows that the exponents of the larger primes in the multiplicative expansion of $P_\alpha - 1$ must obey the constraint

$$\alpha_2 \geq, \alpha_3 \geq, \alpha_5 \geq, \alpha_7, \ldots \qquad (7)$$

in conformance with the factor structure shown in Eq.(6). Overall, this procedure yields an integer for $P_\alpha - 1$ that possesses a number of divisors [26] given by

$$d(P_\alpha - 1) \cong 2.3 \times 10^{11}, \qquad (8)$$

a magnitude only a factor of ~ 3 below the maximum possible [44] for an integer of the size of $P_\alpha$ without compliance with the restriction stemming from the Bézout identity stated in Table II. Hence, although $P_\alpha - 1$ is not a highly composite number, it fulfills the criterion for classification as an abundant number [45,46], since the sum of its divisors exceeds $2(P_\alpha - 1)$. Furthermore, by direct numerical test, this choice of $P_\alpha - 1$ gives both $P_\alpha$ and $2P_\alpha - 1$ as primes of the form $P_\alpha \equiv 1 \pmod{4}$, two crucial constraints in the analysis detailed above. Therefore, this theoretical approach immediately faces three independent and lethal modes of failure; specifically, if $P_\alpha$ is composite, if $P_\alpha$ is a prime congruent to $3 \pmod{4}$, and if $2P_\alpha - 1$ is not prime. The optimized choice for $P_\alpha - 1$ presented in Eq.(6) successfully passes this triply fatal barrier.

The bar to acceptance concerning the joint primality of $P_\alpha$ and $2P_\alpha - 1$ is effective in reducing the availability of alternative choices for $P_\alpha$, since the density of large magnitude low z-smooth integers is very low [30,47] and the density of highly composite numbers [41-43] is sensationally small [44]. We now illustrate the efficacy of the $P_\alpha, 2P_\alpha - 1$ primality screen for highly composite numbers, since, with the neglect of the constraint associated with the Bézout identity stated in Table II, these integers would represent the maximal complexity for the group structure. In the number range spanning $\sim P_\alpha / 10$ to $\sim 10 P_\alpha$, a region greatly exceeding the physically allowed zone [2], there exist only 33 highly composite numbers, all even. With the addition of unity to produce odd parity, none passes the joint $P_\alpha, 2P_\alpha - 1$ primality test. Furthermore, all possess $\alpha_2 \geq 8$ and are consequently far less constrained by Eq.(7) than the form given for $P_\alpha - 1$ by Eq.(6) that is legislated by the condition $\alpha_2 = 2$. Indeed, an examination of the distribution of highly composite numbers [44] reveals that the constraint of $\alpha_2 = 2$ has a profound influence; the largest highly composite number with $\alpha_2 = 2$ is 1260.

Stemming from the composition of $P_\alpha - 1$ given by Eq. (6) and the stipulation stated by Eq.(7), other alternative divisor structures can nevertheless be imagined. Consider the transformation of $P_\alpha - 1$ into an integer $\Gamma$ of the form

$$\Gamma = (7 \cdot 11 \cdot 13 \cdot 17)(P_\alpha - 1) / (149 \cdot 151) \cong 5.11 \times 10^{60}, \tag{9}$$

a number that contains quadratic powers of the primes 7,11,13, and 17, deletes the primes 149 and 151, and has a number of divisors $d(\Gamma) \cong 2.9 \times 10^{11}$, a value modestly exceeding the magnitude of $d(P_\alpha - 1)$ given by Eq.(8). Although $\Gamma + 1$ is not prime, thereby eliminating $\Gamma$ as a specific candidate, we wish to explore the general influence on the divisor structure of $\Gamma$ produced by this form of modification of the prime factors. In spite of the <u>increase</u> in the total inventory of divisors given by $d(\Gamma)$, the resulting number of candidates for the divisors $g_\alpha$ and $g_\beta^{-1}$ that (a) satisfy the condition governed by the Bézout identity given in Table II and (b) potentially give from Eq.(4) the measured value of $\alpha$ suffers <u>a net reduction</u> of a factor of 4. This reversal occurs, since the partition of $P_\alpha - 1$ into $g_\alpha$ and $g_\beta^{-1}$ must preserve the four additional quadratic primes, as <u>individual</u> factors, a consideration echoed in Table V below. Otherwise, we would perforce obtain $\gcd(g_\alpha, g_\beta^{-1}) > 2$, an outcome explicitly violating the Bézout condition. The continuation of this process by the further introduction of quadratic powers to a greater number of small primes in $P_\alpha - 1$ in a manner respecting Eq. (7) simply aggravates this loss of candidates for $g_\alpha$ and $g_\beta^{-1}$. Hence, this strategy for the enhancement of the number of divisors of $P_\alpha - 1$ is rejected.

On the basis of these considerations, we conclude that under the combined physical and mathematical constraints of $P_\alpha \cong 6.7 \times 10^{60}$ and $\alpha_2 = 2$, the integer represented by Eq.(6) for $P_\alpha - 1$ is optimal and yields the maximal number of divisors.

Hence, $P_\alpha$ is unique and the procedure described above has appropriately produced the desired $(P_\alpha, P_\alpha - 1, 2P_\alpha - 1)$ integer triplet in the experimentally designated interval.

Moreover, the sharpness of the recent high-precision measurement [8] of the fine-structure constant α and the additional constraint expressed by the Bézout identity, stated in Table II concerning the mass numbers $g_\alpha$ and $g_\beta^{-1}$ respectively associated with the electron and muon neutrino states illustrated in Fig.(1), impose two further restrictions on the specific divisors of $P_\alpha - 1$ in connection with Eq.(4) that are described in Table IV. Towit, the total number of divisors of the integer $P_\alpha - 1$ is $d(P_\alpha - 1) \cong 2.32 \times 10^{11}$ from Eq.(8); with the number of odd divisors set at $\sim 7.73 \times 10^{10}$, then the number of even divisors is $\sim 1.55 \times 10^{11}$. Since evaluation of the Euler function [26,27,30] in this case gives ~0.1, the expected number of primitive roots in this set of even divisors is $\sim 1.55 \times 10^{10}$. With the interpretation that the $\nu_e$ and $\nu_\mu$ neutrinos represent particles that together carry the full genetic divisor structure of $P_\alpha - 1$ in accord with Eq.(2), but are organized physically to be at maximal genetic separation [12,13] with $D(g_\alpha, g_\beta^{-1}) = 1/2$, by inspection, from the requirement that $\gcd(g_\alpha, g_\beta^{-1}) = 2$, the Bézout identity relates the four states shown in Fig.(1) with the important additive relationship [2]

$$[g_\alpha]_{P_\alpha} [g_\alpha]_{P_\alpha}^{-1} + [g_\beta]_{P_\alpha} [g_\beta]_{P_\alpha}^{-1} \equiv [D(g_\alpha, g_\beta^{-1})]^{-1} \pmod{P_\alpha} \equiv 2 \pmod{P_\alpha} \qquad (10)$$

This statement involves four primitive roots of $P_\alpha$ and supplements the multiplicative seesaw congruence governing $g_\alpha$ and $g_\beta^{-1}$ given by Eq.(2).

**Table V: Test of Divisors of $P_\alpha - 1$ Based on the Precision Measurement of $\alpha$ and the Bézout Identity Restriction**

| Mathematical Property | General Relation | Density Reduction at $P_\alpha$ Number Scale | Remarks | References |
|---|---|---|---|---|
| Divisor Structure of $P_\alpha - 1, \{d_{P_\alpha - 1}\}$ | $P_\alpha - 1 = \prod_{i=1}^{36} P_i^{\alpha_i}$ <br> $d(P_\alpha - 1) = \prod_{i=1}^{36}(\alpha_i + 1)$ | $d(P_\alpha - 1) \cong 2.32 \times 10^{11}$ all divisors, <br> $d((P_\alpha - 1)/4) \cong 7.73 \times 10^{10}$ odd divisors, even divisors $\cong 1.55 \times 10^{11}$. | Primitive roots of $P_\alpha$ that are divisors of $P_\alpha - 1$ with $P_\alpha \equiv 1 \pmod 4$ are perforce even integers. Euler function ~0.10; number of primitive roots among even divisors of $P_\alpha - 1$ is $n_p \cong 1.55 \times 10^{10}$. | 26,30 |
| Fine – Structure Constant $\alpha$ | $g_\alpha g_\beta^{-1} = P_\alpha - 1,$ <br> $\alpha^{-1} = \dfrac{4 g_\beta^{-1}}{g_\alpha}$ | Divisor fractional change is $\delta \sim \tfrac{1}{2}$ measured fractional change in $\alpha (\sim 185\,ppt)$, hence, $\delta \cong 1.85 \times 10^{-10}$. | Globally estimated average number of primitive root divisors of $P_\alpha - 1$ allowed within $\alpha$ measurement range, $N_d \cong n_p \delta \cong 2.9$. <br> Explicit local examination of primitive root divisors of $P_\alpha -1$, restricted to the physically allowed value for $\alpha_2$ given by the width $\Delta$ in Fig.(2), demonstrated that the true local density is ~6-fold less ($\eta \cong 0.16$) than the globally estimated average value. Hence, $N_\eta \cong \eta N_d \cong \eta n_p \delta \cong 0.5$. | 8,26,30 |
| Bézout Identity | $\gcd(g_\alpha, g_\beta^{-1}) = 2$ | $\Omega_B = 4/9$ | Divisors $3^2$ and $5^2$ of $P_\alpha - 1$ segregate in $g_\alpha$ and $g_\beta^{-1}$ as single factors. Condition erects partition in $\mathbb{F}_{P_\alpha}^*$. See Eqs. (10), (12) and (13). | 26,30 |

Summary of Tests Based on Divisors of $P_\alpha$-1

**Number of Expected Candidates for $\alpha^{-1}$ from Eq.(4) is $N_\alpha = \Omega_B N_\eta = \Omega_B \eta n_p \delta \cong 0.2$**

The outcome derived in Table V from the <u>globally</u> estimated <u>average</u> density of primitive roots characteristic of the divisors of $P_\alpha - 1$, together with the restriction specified in Eq. (10) by the Bézout identity on the divisors of $g_\alpha$ and $g_\beta^{-1}$ that leads to the factor $\Omega_B$ in Table V, is the expectation of ~1.3 divisor pairs yielding $\alpha^{-1}$ from Eq. (4) in the range allowed by measurement. This conclusion, based on the global estimate of the number of primitive roots among the divisors, was refined by a direct computationally exhaustive determination of the distribution of the primitive roots associated with the relatively narrow experimental region designated by Δ in Fig. (2). The analysis demonstrated that the true <u>local</u> density of primitive roots is actually ~6-fold less ($\eta \cong 0.16$) than the average given by the global appraisal. With this modification, the final expected number $N_\alpha$ of suitable divisor pairs for $\alpha^{-1}$ corresponding to the region defined by the high-precision data [8] developed in Table V becomes $N_\alpha \cong 0.2$, a value sufficiently less than unity that it favors none.

## EVALUATION OF α

The analysis presented above concludes (a) that $P_\alpha$ is the unique prime defined by the existing physical data and mathematical constraints and (b) that the number of expected primitive root divisor pairs $(g_\alpha, g_\beta^{-1})$ of $P_\alpha - 1$ yielding a value for $\alpha$ that lies within the uncertainty of the high-precision measurement [8] of $\alpha$, if any, is congruent with the existence of a single pair. This projection faithfully fits the specific result illustrated in Fig. (2) that shows nearly exact agreement of the predicted value $\alpha^{-1}$ with the centroid of measured zone giving the experimental magnitude [8] of $\alpha^{-1}$. Quantitatively, the difference between the two is ~146 ppt. Furthermore, since the cryptographic analysis automatically generates an ultra-precision value for $\alpha^{-1}$, stated herein at a precision greater than 1 part in $10^{20}$, it provides a value for $\alpha$ that exceeds all projections [51] of the future capability to measure its magnitude. Since these results are also <u>in exact alignment with a precision cryptographic computation [7] of the cosmological constant $\Omega_\Lambda$,</u> a sensitive <u>independent test</u> of the magnitude of $P_\alpha$, a coherent synthesis in full accord with the corresponding observational data that quantitatively relates the six intrinsic universal parameters α, G, h, c, $\Omega_\Lambda$, $\Omega_m$ and predicts perfect flatness ($\Omega_\Lambda + \Omega_m = 1.0$) is obtained [7].

The specific integers corresponding to the primitive roots $g_\alpha$ and $g_\beta^{-1}$, that yield from Eq.(4) the theoretical value for the fine-structure constant

$$\alpha^{-1} = \frac{4g_\beta^{-1}}{g_\alpha} = 137.0359991047437444154 \tag{11}$$

presented in Fig. (2), are

$$\begin{aligned}g_\alpha &= \\ &2 \cdot 3^2 \cdot 5^2 \cdot 7 \cdot 11 \cdot 17 \cdot 19 \cdot 31 \cdot 47 \cdot 53 \cdot 59 \cdot 61 \cdot 73 \cdot 79 \cdot 103 \cdot 109 \cdot 113 \cdot 131 \cdot 149 \\ &\cong 4.44 \times 10^{29}\end{aligned} \tag{12}$$

and

$$g_\beta^{-1} = 2 \cdot 13 \cdot 23 \cdot 29 \cdot 37 \cdot 41 \cdot 43 \cdot 67 \cdot 71 \cdot 83 \cdot 89 \cdot 97 \cdot 101 \cdot 107 \cdot 127 \cdot 137 \cdot 139 \cdot 151 \cong$$
$$1.52 \times 10^{31}. \qquad (13)$$

With the energy unit $E_0$ specified in Table II, the physical $\nu_e$ and $\nu_\mu$ neutrino masses corresponding to these mass numbers are

$$m_{\nu_e} = g_\alpha E_0 \cong 0.8019 \, \text{meV} \qquad (14)$$

and

$$m_{\nu_\mu} = g_\beta^{-1} E_0 \cong 27.45 \, \text{meV}. \qquad (15)$$

These magnitudes for the $\nu_e$ and $\nu_\mu$ neutrino masses are consistent with existing experimentally determined limits [2,52-54]. Furthermore, since the Higgs state is of order 4 and $g_\alpha$ is a primitive root of $P_\alpha$, we can immediately write [28]

$$g_\alpha^{\frac{P_\alpha - 1}{4}} \equiv B_{\text{Higgs}} (\text{mod } P_\alpha). \qquad (16)$$

This statement is valid for any primitive root of $P_\alpha$ and yields, in conjunction with Eq.(1), the masses of the Higgs supersymmetric pair presented in Table II. Finally, from Table II, we obtain the unified strong-electroweak coupling constant $\alpha^*$ in the form

$$(\alpha^*)^{-1} = \frac{g_\beta^{-1}}{g_\alpha} = \frac{m_{\nu_\mu}}{m_{\nu_e}} = \frac{\alpha^{-1}}{4} \cong 34.26, \qquad (17)$$

a value that is given solely by the $\nu_\mu / \nu_e$ fermion mass ratio and stands in good agreement with its predicted range [2,28,31].

## CONCLUSIONS

The analysis has shown that the description of particle states and interactions can be reduced to a code-breaking problem that is formulated with a <u>physically anchored</u> finite field. The chief result is the ability to provide a theoretical basis for the computation of the magnitude for α that is in full agreement with the measured value, now established at a precision of ~ 370 ppt. Also produced are the predicted magnitudes of the masses of the electron neutrino ($\nu_e$), the muon neutrino($\nu_\mu$), and a supersymmetric Higgs pair that represents a symmetry point of the cryptographic system. A leading physical insight is the finding that α is proportional to the ratio of the electron and muon neutrino masses; the two lightest fermions known establish the strength of this basic coupling. An unexpected outcome is the demonstration that the cryptographic machinery has the ability to amplify the precision of measured results. Although α was measured to approximately one part in $10^{10}$, the resulting predictions of the neutrino masses are given to approximately one part in $10^{30}$ and α becomes known to more than 200 digits. This occurs because the intrinsic precision of the cryptographic system is unity at all scales. Viewed in terms of the operation of measurement, it is equivalent to the complete elimination of noise. The result is the confrontation of an epistemological limit, since the full precision of the theoretical values obtained is surely far beyond any possible human technology of measurement.

## Acknowledgements

Yang Dai is recognized for many key contributions in the earlier phase of the work, specifically, for the important proof of Lemma 1, a set of very complex computations, and the solution of the Higgs congruence. James W. Longworth is thanked for his thorough reading of the manuscript and countless discussions of the ideas and their significance to physical processes. Alex Borisov is acknowledged for many specific numerical calculations. Kevin Ford is acknowledged for discussions concerning highly composite numbers. C. Martin Stickley is thanked for several discussions concerning the basic physical mechanisms and their potential applications. John McCorkindale is acknowledged for his careful calculation of arithmetic functions.

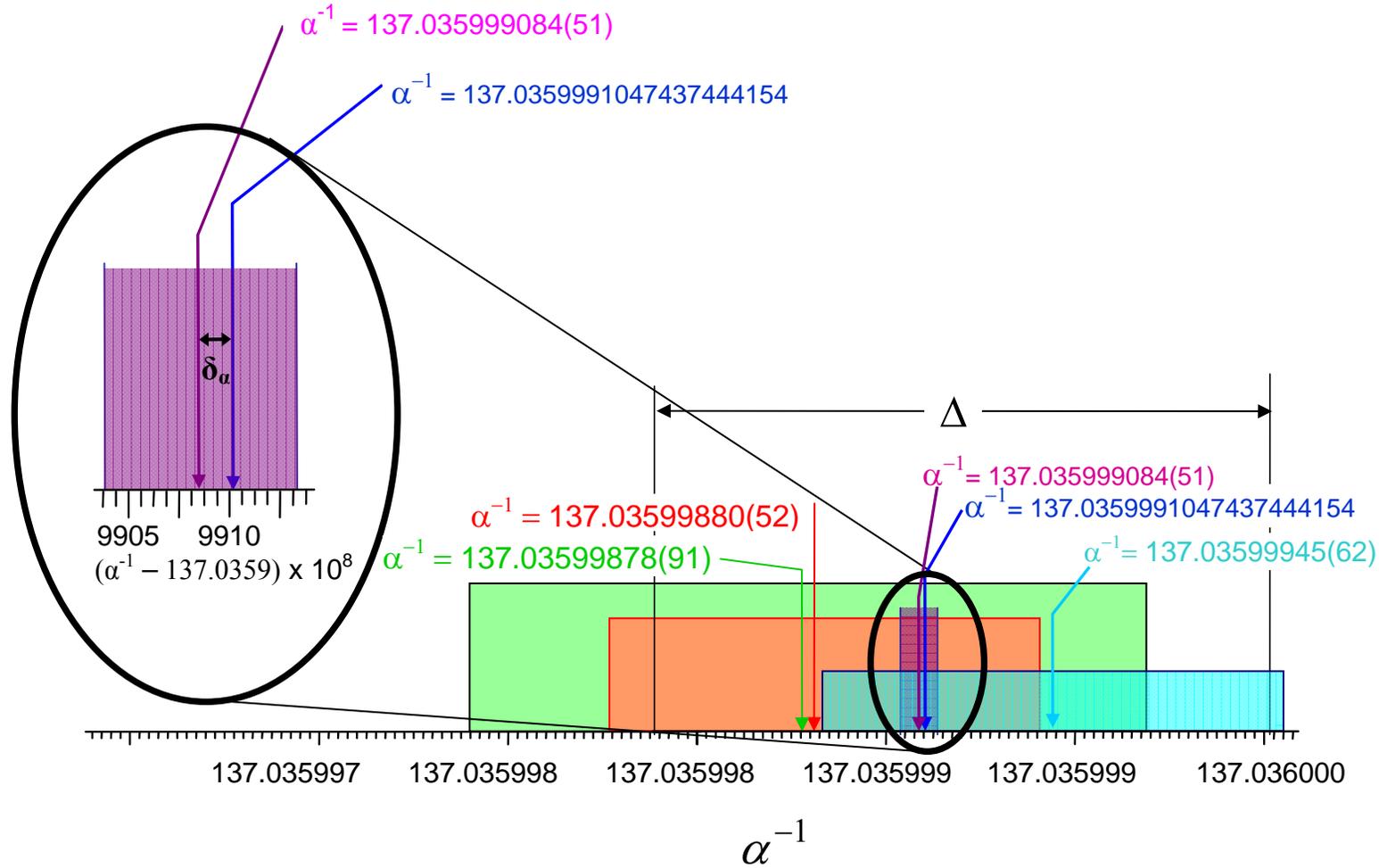

Fig (2): Comparison of selected measured values of the fine-structure constant $\alpha$ with the theoretically predicted magnitude given by Eq.(11). $\Delta$ range, Ref.[48]; orange zone, Ref.[49]; green zone, Ref.[50]; magenta zone, representing the high-precision data giving $\alpha^{-1} = 137.035999084(51)$, Ref.[8]; the cryptographic prediction yielding $\alpha^{-1} = 137.0359991047437444154$ from Eqs.(4) and (11) with $g_\alpha$ and $g_\beta^{-1}$ given respectively by Eqs.(12) and (13). The inset displays the parameter $\delta_\alpha$; with $\delta_\alpha/\alpha^{-1} \cong 146$ ppt, the level of theoretical agreement with the high-precision data is quantified. The theoretical value of $\alpha^{-1}$ specified by Eq.(11) shown is also in agreement with the less precise experimentally measured value $\alpha^{-1} = 137.03599945(62)$ determined by the combination of atomic interferometry with Bloch oscillations shown in light blue, Ref.[9].